# 4D Polytopes and Their Dual Polytopes of the Coxeter Group $W(A_4)$ Represented by Quaternions

**Mehmet Koca** [a)] **, Nazife Ozdes Koca** [b, *)] **and Mudhahir Al-Ajmi** [c)]

Department of Physics, College of Science, Sultan Qaboos University
P. O. Box 36, Al-Khoud 123, Muscat, Sultanate of Oman

**ABSTRACT**

4-dimensional $A_4$ polytopes and their dual polytopes have been constructed as the orbits of the Coxeter-Weyl group $W(A_4)$ where the group elements and the vertices of the polytopes are represented by quaternions. Projection of an arbitrary $W(A_4)$ orbit into three dimensions is made using the subgroup $W(A_3)$. A generalization of the Catalan solids for 3D polyhedra has been developed and dual polytopes of the uniform $A_4$ polytopes have been constructed.

[a)] electronic-mail: kocam@squ.edu.om
[b, *)] electronic-mail: nazife@squ.edu.om, Corresponding author
[c)] electronic-mail: mudhahir@squ.edu.om

## 1. Introduction

The orbits of the Coxeter groups in an arbitrary Euclidean space represent the quasi regular polytopes (including the regular and semi regular polytopes as special cases). An infinite number of rank-2 Coxeter groups which are isomorphic to the dihedral groups $D_n$ exist in 2D dimensions describing the symmetries of the vertex and edge transitive polygons. In 3D, besides the prismatic groups isomorphic to $D_n \times C_2$, there exist three rank-3 Coxeter groups $W(A_3)$, $W(B_3)$ and $W(H_3)$ describing the symmetries of the tetrahedron, octahedron-cube and icosahedron-dodecahedron and their Archimedean polyhedra. Extension of the rank-3 Coxeter groups to rank-4 Coxeter groups is somewhat straightforward. They are the Coxeter groups $W(A_4)$, $W(B_4)$ and $W(H_4)$ with an exception, the group $W(F_4)$. There exist six Platonic solids which are respectively 5-cell (a generalization of tetrahedron), hyperoctahedron-hypercube, 600-cell-120-cell (a generalization of icosahedron-dodecahedron to 4D) and 24-cell respectively which have no correspondence in 3D. In the higher dimensional Euclidean spaces with $D \geq 5$ there exist no correspondences to the groups $W(H_4)$ and $W(F_4)$. But the generalizations of the tetrahedral and octahedral symmetries to higher dimensions are represented by the Coxeter groups $W(A_n)$ and $W(B_n)$ with $n \geq 5$ respectively which describe the $n+1-\text{cell}$ and the hyperoctahedron-hypercube [1].

The rank-4 Coxeter groups can be constructed with the use of finite subgroups of quaternions [2] which make the 4D polytopes more interesting. In this paper we will construct the Coxeter-Weyl group $W(A_4)$ in terms of quaternions which requires the use of two different quaternionic representations of the binary icosahedral group. The Coxeter-Weyl group $W(A_4)$ is the skeleton of the $SU(5)$ Lie group which has been proposed as the Grand Unified Theory in Particle Physics [3]. Constructions of the irreducible representations of the Lie group $SU(5)$ follow the same technique of determination of the orbits of $W(A_4)$. Therefore two techniques are intimately interrelated to each other where the irreducible representations of the Lie Group $SU(5)$ may require several orbits with certain multiplicities.

Our main interest here is the construction of the vertices of the $A_4$ polytopes and their dual polytopes in terms of quaternions. In Section 2 we demonstrate the construction of the Coxeter group $W(A_4)$ in terms of quaternions which follows from the quaternionic roots of the Coxeter-Dynkin diagram of $A_4$. In Section 3 we develop a technique where one decomposes the orbits of $W(A_4)$ into the orbits of $W(A_3)$. This allows us to view the $A_4$ polytopes in three dimensions as polyhedra displaying the tetrahedral symmetry. Section 4 deals with the constructions of the duals of the polytopes discussed in Section 3, a generalization of the Catalan solids to the polytopes in 4D. We discuss our results in Section 5 regarding their use in physics.

## 2. The Coxeter-Weyl Group $W(A_4)$ constructed with quaternions

Let $q = q_0 + q_i e_i$, $(i = 1,2,3)$ be a real unit quaternion with its conjugate defined by $\bar{q} = q_0 - q_i e_i$ and the norm $q\bar{q} = \bar{q}q = 1$. Here the quaternionic imaginary units satisfy the relations

$$e_i e_j = -\delta_{ij} + \varepsilon_{ijk} e_k, \quad (i, j, k = 1,2,3) \tag{1}$$

where $\delta_{ij}$ and $\varepsilon_{ijk}$ are the Kronecker and Levi-Civita symbols and summation over the repeated indices is implicit. With the definition of the scalar product

$$(p, q) = \frac{1}{2}(\bar{p}q + \bar{q}p), \tag{2}$$

quaternions generate the four-dimensional Euclidean space. The group of quaternions is isomorphic to the Lie group $SU(2)$ which is a double cover of the proper rotation group $SO(3)$. Quaternions have an infinite number of cyclic and dicyclic groups in addition to the binary tetrahedral group *T*, binary octahedral group *O* and the binary icosahedral group *I* [4]. The sets of quaternions representing the binary tetrahedral, binary octahedral and the binary icosahedral groups are given as follows. The quaternions of binary tetrahedral group are represented by the set of unit quaternions in (3)

$$T = \{\pm 1, \pm e_1, \pm e_2, \pm e_3, \frac{1}{2}(\pm 1 \pm e_1 \pm e_2 \pm e_3)\}. \tag{3}$$

They also represent the vertices of the 24-cell which can also be represented by the set of quaternions in (4) [5],

$$T' = \{\tfrac{1}{\sqrt{2}}(\pm 1 \pm e_1), \tfrac{1}{\sqrt{2}}(\pm e_2 \pm e_3), \tfrac{1}{\sqrt{2}}(\pm 1 \pm e_2), \tfrac{1}{\sqrt{2}}(\pm e_3 \pm e_1), \tfrac{1}{\sqrt{2}}(\pm 1 \pm e_3), \tfrac{1}{\sqrt{2}}(\pm e_1 \pm e_2)\}. \tag{4}$$

The union of the sets in (3) and (4) represents the binary octahedral group,

$$O = \{T \oplus T'\} \quad . \tag{5}$$

The set of quaternions representing the binary icosahedral group $I$ is given by

$$I = \{T \oplus S\} \tag{6}$$

where the set $S$ is represented by

$$S=\{\frac{1}{2}(\pm\tau\pm e_1\pm\sigma e_3),\frac{1}{2}(\pm\tau\pm e_2\pm\sigma e_1),\frac{1}{2}(\pm\tau\pm e_3\pm\sigma e_2),$$
$$\frac{1}{2}(\pm\sigma\pm e_1\pm\tau e_2),\frac{1}{2}(\pm\sigma\pm e_2\pm\tau e_3),\frac{1}{2}(\pm\sigma\pm e_3\pm\tau e_1),$$
$$\frac{1}{2}(\pm 1\pm\tau e_1\pm\sigma e_2),\frac{1}{2}(\pm 1\pm\tau e_2\pm\sigma e_3),\frac{1}{2}(\pm 1\pm\tau e_3\pm\sigma e_1), \quad (7)$$
$$\frac{1}{2}(\pm\sigma e_1\pm\tau e_2\pm e_3),\frac{1}{2}(\pm\sigma e_2\pm\tau e_3\pm e_1),\frac{1}{2}(\pm\sigma e_3\pm\tau e_1\pm e_2)\}.$$

Here we define $\tau=\frac{1+\sqrt{5}}{2}$ and $\sigma=\frac{1-\sqrt{5}}{2}$ which satisfy the relations $\tau+\sigma=1,\ \tau\sigma=-1$. It is interesting to note that the set of quaternions $I$ represents the vertices of 600-cell [6] and the set $S$ represents the vertices of the snub 24-cell [7].

An orthogonal rotation in 4D Euclidean space can be represented by the group elements of $O(4)$ as [8]

$$[a,b]:q\rightarrow q'=aqb\ ,\ [c,d]^*:q\rightarrow q''=c\bar{q}d\ . \quad (8)$$

When the quaternions $p$ and $q$ take values from the binary tetrahedral group $T$, that is, if $p,q\in T$ the set of elements of the group

$$(W(D_4)/C_2):S_3=\{[p,q]\oplus[p,q]^*\}\ \ ;\ p,q\in T \quad (9)$$

represents the symmetry of the snub 24-cell [7] of order 576. Similarly if $p,q\in O$ then the set of elements

$$Aut(F_4)\approx W(F_4):C_2=\{[p,q]\oplus[p,q]^*\}\,;\,p,q\in O \quad (10)$$

represents the automorphism group of $F_4$ [9]. If $p,q\in I$ then we have [10]

$$W(H_4)=\{[p,q]\oplus[p,q]^*\}. \quad (11)$$

All these correspondences between the Coxeter groups and the finite subgroups of quaternions are interesting. Some further discussions can be found in the reference [11].

Now we construct the group elements of the Coxeter-Weyl group $W(A_4)$ in terms of quaternions. The Coxeter-Dynkin diagram of $A_4$ is given in Fig.1 where the simple roots scaled by $\sqrt{2}$ can be taken as

$$\alpha_1=-1,\quad \alpha_2=\frac{1}{2}(1+e_1+e_2+e_3),\quad \alpha_3=-e_1,\quad \alpha_4=\frac{1}{2}(e_1-\sigma e_2-\tau e_3). \quad (12)$$

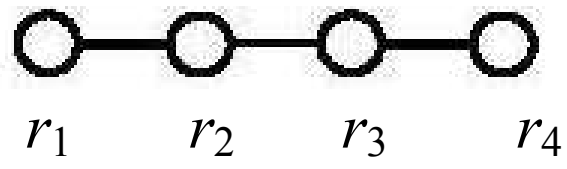

**Fig. 1.** The Coxeter diagram of $A_4$.

The generators $r_i = [\alpha_i, -\alpha_i]^*$, ( $i = 1,2,3,4$) generate the Coxeter–Weyl group $W(A_4)$ which can be compactly written as follows

$$W(A_4) = \{[p, \bar{c}\bar{\tilde{p}}c] \oplus [p, c\bar{\tilde{p}}c]^*\}, \tag{13}$$

or symbolically, $W(A_4) = \{[I, \bar{c}\bar{\tilde{I}}c] \oplus [I, c\bar{\tilde{I}}c]^*\}$.

Here $p \in I$ is an arbitrary element of the binary icosahedral group $I$ with $c = \frac{1}{\sqrt{2}}(e_3 - e_2) \in T'$ and $\tilde{p} = p(\tau \leftrightarrow \sigma)$ is an element of the binary icosahedral group $\tilde{I}$ obtained from $I$ by interchanging $\tau$ and $\sigma$. The Dynkin diagram symmetry leads to the group extension, namely,

$$Aut(A_4) = W(A_4) : \gamma = \{[\,p, \pm\bar{c}\bar{\tilde{p}}c\,] \oplus [\,p, \pm c\bar{\tilde{p}}c\,]^*\} . \tag{14}$$

The extended Coxeter-Dynkin diagram is shown in Fig.2 where the generator $r_0 = [\alpha_0, -\alpha_o]^*$ is the reflection with respect to the hyperplane orthogonal to the root

$\alpha_0 = -(\alpha_1 + \alpha_2 + \alpha_3 + \alpha_4) = \frac{1}{2}(1 - \tau\, e_2 - \sigma\, e_3)$.

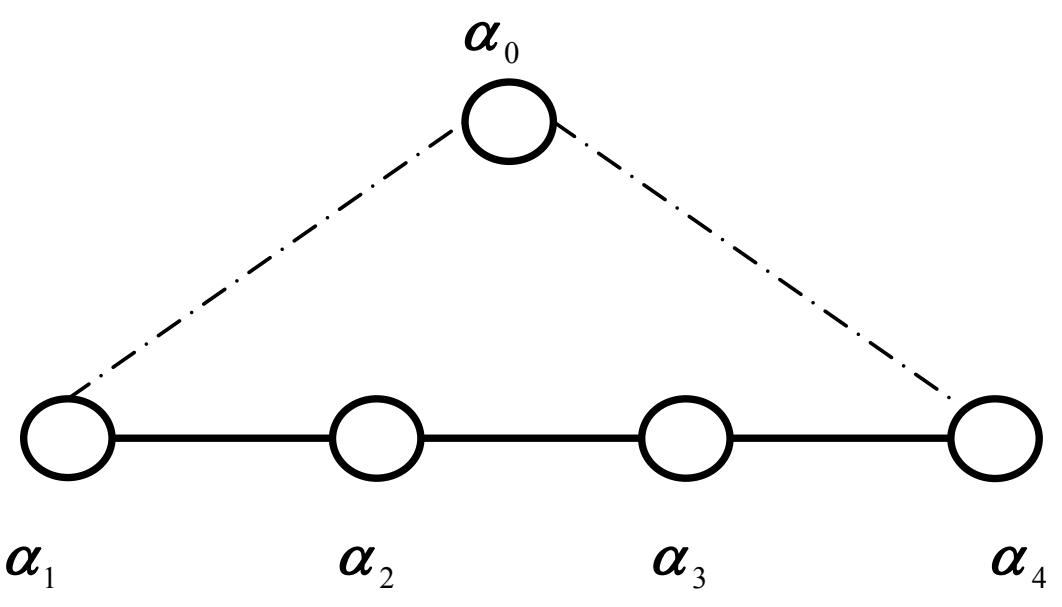

**Fig. 2.** The extended diagram of $A_4$.

The group $W(A_4)$ has three maximal subgroups up to conjugations, the tetrahedral group $W(A_3) \approx S_4$ of order 24, the group $W(A_2 \times A_1) \approx D_3 \times C_2$ of order 12 and the dihedral group $W(H_2) \approx D_5$ of order 10. The group $W(A_3) \approx S_4$ can be embedded in the group $W(A_4) \approx S_5$ five different ways which can be represented by the sets of generators $(r_1, r_2, r_3), (r_2, r_3, r_4), (r_3, r_4, r_0), (r_4, r_0, r_1), (r_0, r_1, r_2)$.

Denote by $R_1 = r_1 r_3$ and $\mathrm{R}_2 = \mathrm{r}_2\mathrm{r}_4$, the generators of the Coxeter group $W(H_2)$ where $d = R_1 R_2 = [\alpha, \beta]$ is the Coxeter element satisfying $d^5 = 1$ with the elements

$\alpha = \frac{1}{2}(-\sigma + e_2 + \tau e_3)$ and $\beta = \frac{1}{2}(-\tau + \sigma e_2 + e_3) = \bar{c}\tilde{\bar{\alpha}}c$.

**3. Construction of the $W(A_4)$ orbits as 4D polytopes**

An arbitrary 'highest' weight vector $\Lambda$ describing an irreducible representation of the Lie group $SU(5)$ is given in the Dynkin basis [12] as $\Lambda = (a_1 a_2 a_3 a_4) \equiv a_1\omega_1 + a_2\omega_2 + a_3\omega_3 + a_4\omega_4$ where $\omega_i, (i = 1,2,3,4)$ are the vectors of the dual basis satisfying the relation $(\sqrt{2}\alpha_i, \omega_j) = \delta_{ij}$. For the 'highest' weight vector the Dynkin indices are non-negative integers $a_i \geq 0$. The Cartan matrix defined by $(\alpha_i, \alpha_j) = \frac{1}{2}C_{ij}$ and its inverse $(\omega_i, \omega_j) = (C^{-1})_{ij}$ are given by

$$C = \begin{pmatrix} 2 & -1 & 0 & 0 \\ -1 & 2 & -1 & 0 \\ 0 & -1 & 2 & -1 \\ 0 & 0 & -1 & 2 \end{pmatrix}, \quad C^{-1} = \frac{1}{5}\begin{pmatrix} 4 & 3 & 2 & 1 \\ 3 & 6 & 4 & 2 \\ 2 & 4 & 6 & 3 \\ 1 & 2 & 3 & 4 \end{pmatrix}. \qquad (15)$$

The dual vectors $\omega_i$ can be written in terms of quaternions as follows

$$\omega_1 = \frac{1}{\sqrt{10}}(-\sqrt{5} + \tau e_2 - \sigma e_3), \qquad \omega_2 = \frac{1}{\sqrt{10}}(2\tau e_2 - 2\sigma e_3),$$
$$\omega_3 = \frac{1}{\sqrt{10}}(-\sqrt{5}e_1 + \tau^2 e_2 - \sigma^2 e_3), \quad \omega_4 = \frac{1}{\sqrt{10}}(2e_2 - 2e_3) \equiv -\frac{2}{\sqrt{5}}c. \tag{16}$$

We define an orbit of the group $W(A_4)$ by the group action $O(\Lambda) \equiv W(A_4)\Lambda$.

Since we are not in general concerned with the orbits of the irreducible representations of the Lie group SU(5), we can take the vector $\Lambda = (a_1 a_2 a_3 a_4)$ with $a_i \geq 0$ any non-negative real number, not particularly restricted to integers only. The size of the orbit $O(\Lambda) = W(A_4)\Lambda$ is 120 with $a_i \neq 0, i = 1,2,3,4$ which is equal to the order of the group $W(A_4) \approx S_5$. One can count the number of cells of the polytope $O(\Lambda) = W(A_4)\Lambda$ by using the maximal subgroups $W(A_3)$ and $W(A_2 \times A_1)$. We will do it later for particular cases where $a_i$ take certain values. The right-coset decomposition of the group $W(A_4)$ under its maximal subgroup $W(A_3)$ can be given by

$$W(A_4) = \{W(A_3) \oplus W(A_3)d \oplus W(A_3)d^2 \oplus W(A_3)d^3 \oplus W(A_3)d^4\} \tag{17}$$

where $d$ can be any Coxeter element of the group $W(A_4)$. Since all the Coxeter elements of the Coxeter groups are conjugate to each other we choose it without loss of generality as the Coxeter element $d = R_1 R_2 = [\alpha, \beta]$. Then the orbit $O(\Lambda) = W(A_4)\Lambda$ decomposes as $O(\Lambda) = \sum_{i=0}^{4} \oplus W(A_3)d^i\Lambda$. Similarly the decomposition of the orbit $O(\Lambda) = W(A_4)\Lambda$ under the group $W(A_2 \times A_1) \approx D_3 \times C_2$ can be written as

$$O(\Lambda) = \sum_{i=0}^{4} \oplus W(A_2 \times A_1)d^i\Lambda \oplus W(A_2 \times A_1)R_1\Lambda .$$

These two decompositions are useful for the projections of the 4D polytopes of the group $W(A_4)$ into three dimensions under the tetrahedral group and the prismatic groups respectively. One can also project the polytope $O(\Lambda) = W(A_4)\Lambda$ into 2D by using the decomposition $W(A_4) = \sum_{i=1}^{12} \oplus W(H_2)h_i, \quad h_i \in W(A_2 \times A_1)$. In this paper we will be interested in the projection under the group $W(A_3)$. Let us denote the orbit of the Coxeter element $d$ by the vectors $\Lambda(i) \equiv d^i\Lambda$. It is evident that the orbit $O(\Lambda) = W(A_4)\Lambda$ is the union of the five orbits $W(A_3)\Lambda(i), \ i = 0,1,2,3,4.$ We choose the subgroup $W(A_3)$ generated by

$$W(A_3) = \langle r_1, r_2, r_3 \rangle = \{[T, \bar{c}\bar{T}c] \oplus [T, c\bar{T}c]^*\} . \tag{18}$$

Here the vector $c$ is orthogonal to the first three scaled roots $\alpha_i$, $i=1,2,3$. This shows that the tetrahedral group $W(A_3)$ leaves the vector $c=-\frac{\sqrt{5}}{2}\omega_4$ invariant. In the dual basis, the five vectors $\Lambda(i)\equiv d^i\Lambda$ can be written as

$$\begin{aligned}
&\Lambda(0)\equiv\Lambda=a_1\omega_1+a_2\omega_2+a_3\omega_3+a_4\omega_4\\
&\Lambda(1)\equiv-(a_1+a_2)\omega_1+(a_1+a_2+a_3+a_4)\omega_2-(a_2+a_3+a_4)\omega_3+(a_2+a_3)\omega_4\\
&\Lambda(2)\equiv-(a_3+a_4)\omega_1+a_3\omega_2-(a_1+a_2+a_3)\omega_3+a_1\omega_4\\
&\Lambda(3)\equiv a_4\omega_1-(a_2+a_3+a_4)\omega_2+a_2\omega_3-(a_1+a_2)\omega_4\\
&\Lambda(4)\equiv(a_2+a_3)\omega_1-(a_1+a_2+a_3)\omega_2+(a_1+a_2+a_3+a_4)\omega_3-(a_3+a_4)\omega_4.
\end{aligned}\tag{19}$$

Any orbit $W(A_3)\Lambda(i),\ i=0,1,2,3,4$ involves in general 24 vectors. Let a vector of the Coxeter group $W(A_3)$ be written as $\lambda=(b_1b_2b_3)\equiv b_1v_1+b_2v_2+b_3v_3$ where we have $(\alpha_i,\alpha_j)=\frac{1}{2}C_{ij}$, $(v_i,v_j)=(C^{-1})_{ij}$, $(\sqrt{2}\alpha_i,v_j)=\delta_{ij}$, $i,j=1,2,3$. The Cartan matrix of the group $W(A_3)$ and its inverse are given by

$$C=\begin{pmatrix}2&-1&0\\-1&2&-1\\0&-1&2\end{pmatrix},\qquad C^{-1}=\frac{1}{4}\begin{pmatrix}3&2&1\\2&4&2\\1&2&3\end{pmatrix}.\tag{20}$$

Using (15) and (20) we can express the basis vectors $v_i$ in terms of the vectors $\omega_i$:

$$v_1=\omega_1-\frac{1}{4}\omega_4,\quad v_2=\omega_2-\frac{1}{2}\omega_4,\quad v_3=\omega_3-\frac{3}{4}\omega_4.\tag{21}$$

In the basis $v_i$ the vectors in (19) can be expressed in terms of $v_1,v_2,v_3$ and $\omega_4$ ; one of them, say, $\Lambda(0)$ can be written as $\Lambda(0)\equiv a_1v_1+a_2v_2+a_3v_3+\frac{1}{4}(a_1+2a_2+3a_3+4a_4)\omega_4$. The orbit $W(A_3)\Lambda(0)$ is quasi regular polyhedra with 24 vertices. Similarly the other vectors in (19) lead to polyhedra with 24 vertices provided $a_i\neq0$. Clearly, one can choose other vectors within a given orbit of $W(A_3)$ where the components of $v_i$ can be chosen positive. Then the vectors $\Lambda(i),\ i=1,2,3,4$ can be replaced by $\Lambda'(i)$ as follows

$$\Lambda'(0)\equiv a_1v_1+a_2v_2+a_3v_3+\frac{1}{4}(a_1+2a_2+3a_3+4a_4)\omega_4$$

$$\Lambda'(1)\equiv a_1v_1+a_2v_2+(a_3+a_4)v_3+\frac{1}{4}(a_1+2a_2+3a_3-a_4)\omega_4$$

$$\Lambda'(2) \equiv a_1 v_1 + (a_2 + a_3) v_2 + a_4 v_3 + \frac{1}{4}(a_1 + 2a_2 - 2a_3 - a_4)\omega_4$$

$$\Lambda'(3) \equiv (a_1 + a_2) v_1 + a_3 v_2 + a_4 v_3 + \frac{1}{4}(a_1 - 3a_2 - 2a_3 - a_4)\omega_4$$

$$\Lambda'(4) \equiv a_2 v_1 + a_3 v_2 + a_4 v_3 + \frac{1}{4}(4a_1 + 3a_2 + 2a_3 + a_4)\omega_4 \,. \tag{22}$$

Now we can write the orbit $O(\Lambda) = W(A_4)\Lambda$ in terms of the orbits $O(\Lambda'(i) = W(A_3)\Lambda'(i),\ i = 0,1,2,3,4$ expressed as

$$\begin{aligned} O(a_1 a_2 a_3 a_4) &= O(a_1 a_2 a_3)(-(a_1 + 2a_2 + 3a_3 + 4a_4)) + O(a_1 a_2 (a_3 + a_4))(-(a_1 + 2a_2 + 3a_3 - a_4)) \\ &+ O(a_1 (a_2 + a_3) a_4)(-(a_1 + 2a_2 - 2a_3 - a_4)) + O((a_1 + a_2) a_3 a_4)(-(a_1 - 3a_2 - 2a_3 - a_4)) \\ &+ O(a_2 a_3 a_4)(4a_1 + 3a_2 + 2a_3 + a_4). \end{aligned} \tag{23}$$

Here we followed the normalization of the reference [12] to associate the numbers in the second parenthesis with the $U(1)$ charge of the $SU(4) \times U(1) \subset SU(5)$ Lie theory of the particle physics. For example, the orbit $O(1111)$ can be decomposed under the orbits of $W(A_3)$ as

$$O(1111) = O(111)(-10) + O(112)(-5) + O(121)(0) + O(211)(5) + O(111)(10)\,. \tag{24}$$

One can use the quaternionic bases in (16) to apply the quaternionic representation of $W(A_3)$ in (18). In order to represent the 4D-polytopes represented by the set of vectors $W(A_3)\Lambda(i)$ in three dimensions we should make another change of bases from the set $(1, e_1, e_2, e_3)$ to the set of quaternions $(p_0 \equiv c,\ p_1 \equiv e_1 p_0,\ p_2 \equiv e_2 p_0,\ p_3 \equiv e_3 p_0)$. This is because $p_0$ is left invariant by the group $W(A_3) = \langle r_1, r_2, r_3 \rangle$. We shall be dealing with the orbit $W(A_3)\Lambda'(0)$ only because the others are treated in a similar manner. Let us also define a new set of parameters in terms of the components of $v_1, v_2, v_3$ by

$$\alpha = \frac{1}{2}(a_1 - a_3),\quad \beta = \frac{1}{2}(a_1 + a_3),\quad \gamma = \frac{1}{2}(a_1 + 2a_2 + a_3)\ . \tag{25}$$

Then, in the new basis of quaternions, the vectors of the orbit $W(A_3)\Lambda'(0)$, apart from a common 4$^{th}$ component of $-\frac{p_0}{2\sqrt{5}}(a_1 + 2a_2 + 3a_3 + 4a_4)$, can be written as[13]

$$\begin{aligned} &\pm \alpha p_1 \pm \beta p_2 \pm \gamma p_3, \quad &&\pm \beta p_1 \pm \gamma p_2 \pm \alpha p_3, \quad &&\pm \gamma p_1 \pm \alpha p_2 \pm \beta p_3, \\ &\pm \alpha p_1 \pm \gamma p_2 \pm \beta p_3, \quad &&\pm \gamma p_1 \pm \beta p_2 \pm \alpha p_3, \quad &&\pm \beta p_1 \pm \alpha p_2 \pm \gamma p_3, \end{aligned} \tag{26}$$

where only even number of $(-)$ sign is taken. The decomposition of the other vectors in the new basis of quaternions can be made in a similar manner by redefining the parameters $\alpha, \beta$ and $\gamma$. Before we proceed further it is imperative to discuss the polyhedra represented by the set of vectors in (26). For example the $W(A_3)$ orbit $O(100)$ is given by the vertices $\frac{1}{2}(\pm p_1 \pm p_2 \pm p_3)$ with an even number of $(-)$ sign which represents a tetrahedron [13]. In the reference [14] we have classified all quasiregular polyhedra and their dual polyhedra possessing tetrahedral, octahedral and icosahedral symmetries.

An orbit $W(A_3)\Lambda = O(a_1 a_2 a_3)$ with $a_1 \neq a_2 \neq a_3 \succ 0$ represents a quasi regular truncated octahedron with 24 vertices with isogonal hexagon faces with two edge lengths $(a_1, a_2)$ and $(a_2, a_3)$ attached to a rectangle with edges $(a_1, a_3)$. Since the norms of the roots of the Coxeter diagrams are assumed to be $\sqrt{2}$ for any Dynkin indices the actual length must be taken as $\sqrt{2}a_i$ which we omit in the following discussions. A special case, e.g., $O(a_1 a_2 0)$ represents a quasiregular truncated tetrahedron with isogonal hexagons of edges $(a_1, a_2)$ and triangular faces with edge $a_2$. When $a_1 = a_2$ then we obtain a truncated tetrahedron with all edges having the same length. The orbit $O(a_1 0 a_3)$ represents a quasi regular cuboctahedron with two types of equilateral triangles having edge lengths $a_1$ and $a_2$ respectively and a rectangle with edge lengths $a_1$ and $a_2$. The orbits $O(010)$ and $O(001)$ represent an octahedron and a tetrahedron respectively. In what follows we discuss the regular uniform $W(A_4)$ polytopes. The names of the following regular uniform polytopes are taken from an article in Wikipedia [15].

### 3.1 The polytope $O(1000)$ (5-cell)

This is known as 5-cell which can be decomposed as $O(1000) = O(100)(-1) + O(0)(4)$. The polytope is made of five tetrahedral cells. In the dual basis its vertices are given by

$$O(1000) = \{\ \omega_1,\ \omega_2 - \omega_1,\ \omega_3 - \omega_2, \omega_4 - \omega_3,\ -\omega_4\ \}. \tag{27}$$

When we apply the subgroup $W(A_3) = \langle r_1, r_2, r_3 \rangle$ on the vectors in (27) it is clear that the vector $-\omega_4$ is left invariant and the others are permuted. Actually, any one of the five vectors in (27) is left invariant under one of the five conjugate groups of the tetrahedral group $W(A_3)$. This proves that any four vectors in (27) form a tetrahedron under one of the conjugate tetrahedral groups implying that five vertices in (27) form a 5-cell. In the basis $(p_0, \mathrm{p}_1, \mathrm{p}_2, \mathrm{p}_3)$ the set of vectors of 5-cell in (27) can be written as

$$-\frac{p_0}{2\sqrt{5}} + \frac{1}{2}(p_1 + p_2 + p_3),\ -\frac{p_0}{2\sqrt{5}} + \frac{1}{2}(p_1 - p_2 - p_3),\ -\frac{p_0}{2\sqrt{5}} + \frac{1}{2}(-p_1 - p_2 + p_3),$$

$$-\frac{p_0}{2\sqrt{5}} + \frac{1}{2}(-p_1 + p_2 - p_3),\ \frac{4p_0}{2\sqrt{5}}. \tag{28}$$

This proves that the first four vectors represent the vertices of a tetrahedron representing the points on a 2-sphere which is obtained as the intersection of the hyperplane orthogonal to the unit vector $p_0$ with the 3-sphere represented by the unit quaternions. The last vector represents a point in 3D space.

### 3.2 The polytope $O(0100)$ (rectified 5-cell)

This orbit has 10 vertices which decomposes as $O(0100) = O(010)(-1) + (100)(4)$ expressed in terms of the quaternions in the basis $(p_0,\ p_1,\ p_2,\ p_3)$

$$
\begin{aligned}
&\mathrm{O}(0100) = \{-\frac{2p_0}{2\sqrt{5}} + (\pm\, p_1,\ \pm\, p_2,\ \pm\, p_3);\\
&\frac{3p_0}{2\sqrt{5}} + \frac{1}{2}(\pm\, p_1 \pm p_2 \pm p_3)\ \text{(even number of (-) sign)}\}.
\end{aligned}
\tag{29}
$$

The first 6 components represent an octahedron and the last 4 components represent a tetrahedron. The vertices of the polytope $O(0010)$ are obtained by taking the negatives of those in (29). Similarly the vertices of the polytope $O(0001)$ are the negatives of the set in (28).

### 3.3 The polytope $O(1100)$ (truncated 5-cell)

$O(1100) = O(110)(-3) + (200)(2) + (100)(7)$

The polytope has 20 vertices which are given by three sets of vectors under the $W(A_3)$ decomposition

$$
\begin{aligned}
O(1100) = \{&-\frac{3p_0}{2\sqrt{5}} + [\frac{1}{2}(\pm 3p_1 \pm p_2 \pm p_3), \frac{1}{2}(\pm 3p_1 \pm p_2 \pm p_3)\frac{1}{2}(\pm 3p_1 \pm p_2 \pm p_3)]\\
&\oplus \frac{2p_0}{2\sqrt{5}} + (\pm p_1 \pm p_2 \pm p_3) \oplus \frac{7p_0}{2\sqrt{5}} + \frac{1}{2}(\pm p_1 \pm p_2 \pm p_3)\}\text{(even number of (-) sign)}\}.
\end{aligned}
\tag{30}
$$

This shows that when projected into 3D space the truncated 5-cell is represented by a truncated tetrahedron and two tetrahedra.

### 3.4 The polytope $O(1010)$ (cantellated 5-cell)

$O(1010) = O(110)(1) + (101)(-4) + (010)(6)$

The vertices can be written as

$$O(1010)=\{\frac{p_0}{2\sqrt{5}}+[\frac{1}{2}(\pm 3p_1\pm p_2\pm p_3),\frac{1}{2}(\pm 3p_1\pm p_2\pm p_3)\frac{1}{2}(\pm 3p_1\pm p_2\pm p_3)$$

(even number of(-) sign)}

$$\oplus\{-\frac{4p_0}{2\sqrt{5}}+(\pm p_1\pm p_2,\pm p_2\pm p_3,\ \pm p_3\pm p_1)\}$$

$$\oplus\{\frac{6p_0}{2\sqrt{5}}+(\pm p_1,\pm p_2,\ \pm p_3)\}. \qquad (31)$$

This indicates that the cantellated 5-cell is viewed in 3D as one truncated tetrahedron, one cuboctahedron and one octahedron. We hope that the explicit examples of the vectors in the basis $(p_0,\ p_1,\ p_2,\ p_3)$ are so clear that we don't need to present the remaining cases in terms of quaternions.

### 3.5 The polytope $O(1001)$ (runcinated 5-cell)

Its cells consist of the tetrahedra and triangular prisms and its 20 vertices can be de composed as

$$O(1001)=O(100)(-5)+(101)(0)+(001)(5) \qquad (32)$$

which represents two tetrahedra and one cuboctahedron.

### 3.6 The polytope $O(0110)$ (bitruncated 5-cell)

Its cells are truncated tetrahedra. Its decomposition is given by

$$O(0110)=O(110)(5)+(020)(0)+(011)(-5).$$

Its projection in 3D yields to two truncated tetrahedra and one octahedron.

### 3.7 The polytope $O(1110)$ (cantitrancated 5-cell)

It is a polytope with 60 vertices. Its cells involve truncated octahedra, truncated tetrahedra and triangular prisms. Its decomposition is given by

$$O(1110)=O(111)(-6)+(120)(-1)+(210)(4)+(110)(9). \qquad (33)$$

The first orbit in the decomposition is a truncated octahedron with 24 vertices; the second is a truncated tetrahedron with 12 vertices where the edges of the triangular faces are of 2 units while the isogonal hexagonal face has two edges of length with 2 units and 1 unit. The third polyhedron is also a truncated tetrahedron with 12 vertices where the edge lengths with respect to the previous case are reversed. The last one is the usual truncated tetrahedron with 12 vertices.

### 3.8 The polytope $O(1101)$ (runcitruncated 5-cell)

This is also a polytope with 60 vertices consisting of the cells of truncated tetrahedron, cuboctahedron, hexagonal prism and triangular prism. The decomposition here is given by

$$O(1101)=O(110)(-7)+(111)(-2)+(201)(3)+(101)(8). \quad (34)$$

In 3D it is viewed as a truncated tetrahedron(12 vertices), a truncated octahedron with (24) vertices and a quasi regular polyhedron with triangular faces of edges 2 unit and 1 unit joined to a rectangular face with edges 2 unit and 1 unit. The last polyhedron is a cuboctahedron.

### 3.9 The polytope $O(1011)$ (runcitruncated 5-cell)

It is obtained from the orbit $O(1101)$ by applying the Dynkin diagram symmetry so that the vertices are obtained from the vertices of $O(1101)$ by changing the signs of the vectors.

### 3.10 The polytope $O(1111)$ (omnitruncated 5-cell)

This polytope with its 120 vertices consists of the cells of truncated octahedra and hexagonal prisms. When it is projected into 3D under the group $W(A_3)$ one obtains the relations given in (24). The polyhedron $O(111)$ is a truncated octahedron which appears twice. The quasi regular polyhedron $O(112)$ consists of the hexagonal faces of edge length 1 unit and isogonal hexagon of two edge lengths with 2 units and 1 unit which is joined to a rectangular face of edges of lengths 2 and 1 unit. The polyhedron $O(211)$ is obtained from the one $O(112)$ by the Dynkin diagram symmetry. The polyhedron $O(121)$ consists of the faces of isogonal hexagons with edges of 2 unit and 1 unit and two of which are connected to a square of side 1 unit.

## 4. Duals of the uniform polytopes of the Coxeter-Weyl group $W(A_4)$

Determination of the Archimedean solids dates back to the years of Archimedes lived in 287BC-212BC. After nearly 20 centuries the dual polyhedra of the Archimedean solids, so called Catalan solids, are determined by Catalan in 1865[16]. To our surprise duals of the 4D uniform polytopes have not been studied so far. In this section we discuss the construction of the duals of the 4D uniform polytopes possessing the $W(A_4)$ symmetry. In the references [17] and [18] we have proved that all the Catalan solids can be derived from the Coxeter diagrams $A_3, B_3, H_3$ with a simple technique which we will also apply for the constructions of the duals of the 4D polytopes. If we take a vector with the Dynkin index $(a_1a_2a_3)$ then all orbits with $a_i=0$ or 1 except $O(000)$ represent the Archimedean solids including the platonic polyhedra. But snub cube and snub dodecahedron needs a different analysis in terms of Coxeter diagrams [18]. It is then quiet natural to extend the same analysis to the rank-4 Coxeter diagrams and classify the uniform polytopes $O(a_1a_2a_3a_4)=W(G)(a_1a_2a_3a_4)$ as the platonic and Archimedean polytopes where $a_i=0$ or 1 and the $G$ is one of the diagrams $A_4, B_4, H_4, F_4$ . However, in mathematical literature, the name of the Archimedean polytope is strictly used for the polytope whose cells are made of platonic solids only. In what follows we will discuss the construction of the duals of the uniform polytopes possessing the Coxeter-Weyl symmetry $W(A_4)$.

To obtain the vertices of the dual polytope of a given polytope one determines the centers of the cells joined to the vertex $(a_1a_2a_3a_4)$. The relative magnitudes of these vectors are determined from the fact that the hyperplane formed by the vectors representing the centers of the cells are orthogonal to the vertex $(a_1a_2a_3a_4)$. Let us recall that the dual polytopes are cell transitive similar to the Catalan solids which are face transitive. In what follows we discuss each $W(A_4)$ uniform polytope in turn.

**4.1 Dual of the 5-cell** $O(1000) = O(\omega_1)$

Let us recall that the Coxeter-Weyl group $W(A_4) = \langle r_1, r_2, r_3, r_4 \rangle$ acting on the vector $\omega_1 = (1000)$ generates the vertices of the 5-cell. The number of vertices of a polytope is determined as the index of the subgroup leaving the vector, in this case $\omega_1$, invariant. Since the group fixing $\omega_1$ is generated by the generators $\langle r_2, r_3, r_4 \rangle$ then the index $\frac{|W(A_4)|}{|W(A_3)|} = 5$ determines the number of vertices of the polytope. Similarly the number of the cells of the polytope is determined by the index of the subgroup $W(A_3) = \langle r_1, r_2, r_3 \rangle$ generating the tetrahedral cell, thus, $\frac{|W(A_4)|}{|W(A_3)|} = 5$ . The center of the first tetrahedron generated by the generators $\langle r_1, r_2, r_3 \rangle$ can be taken as $\omega_4 = (0001)$, up to a scale factor, since it is left invariant by the generators $\langle r_1, r_2, r_3 \rangle$ . Then one can apply the group $W(A_4)\omega_4$ to obtain 5 vectors which constitute another 5-cell. Indeed one notes that $W(A_4)\omega_4 = -W(A_4)\omega_1$. It is this fact that 5-cell is said to be self- dual.

**4.2 Dual of the rectified 5-cell** $O(0100) = O(\omega_2)$

This uniform polytope is an Archimedean polytope since its cells are made of octahedra and tetrahedra. The vector $\omega_2$ is left invariant under the group $W(A_2 \times A_1) = \langle r_1, r_3, r_4 \rangle$.Therefore the number of vertices is given by $\frac{|W(A_4)|}{|W(A_2 \times A_1)|} = 10$. Of the two types of cells, the octahedral cells are generated by the tetrahedral group $W(A_3) = \langle r_1, r_2, r_3 \rangle$ along with its five conjugate groups and the tetrahedral cells are generated by the group $W(A_3) = \langle r_2, r_3, r_4 \rangle$ and its conjugates. Therefore the rectified 5-cell has 10 cells. The centers of the cells generated by these two groups upon the action on the vector $\omega_2$ can be taken to be $\omega_4$ and $\omega_1$ respectively. One can also determine the number of cells which involves $\omega_2$ as one of the vertices. The number of octahedral cells attached to the vector $\omega_2$ is given by the formula

$$\frac{|W(A_4)|}{|W(A_3)|} \times \frac{\text{number of vertices of the octahedron}}{\text{number of vertices of the polytope}} = \frac{120\times 6}{24\times 10} = 3 .$$

Similarly the number of tetrahedral cells joined to the vector $\omega_2$ is equal to 2. One can easily check that the centers of the three octahedral cells are given by the vectors

$$\omega_4,\ r_3r_4\omega_4,\ (r_3r_4)^2\omega_4 = r_4r_3\omega_4 \tag{35}$$

which form an equilateral triangle. The centers of the tetrahedral cells joined to $\omega_2$ are given by $\omega_1$ and $r_1\omega_1 = (\omega_1 - \alpha_1)$ up to a scale factor. We note that the vector $\omega_1$ - $r_1\omega_1 = \alpha_1$ is orthogonal to the plane determined by three vectors in (35). These five vertices must be in the hyperplane orthogonal to the vector $\omega_2$. Then the scale factor can be determined from the scalar product $(\lambda\omega_1 - \omega_4).\omega_2 = 0$ leading to $\lambda = \frac{2}{3}$. One can show that any line joining these five vertices lie in the hyperplane orthogonal to the vertex $\omega_2$. The dual polytope has 10 vertices consisting of the union of the orbits $O(1000) \oplus \frac{2}{3}\mathrm{O}(0001)$ with 10 cells. Each cell of the dual polytope is a dipyramid of edge lengths $\sqrt{2}$ and $\frac{3}{\sqrt{2}}$. The vertices of the dual polytope are on two concentric 3-spheres with radii $\frac{2}{\sqrt{5}}$ and $\frac{3}{\sqrt{5}}$. The coordinates of this cell in 3D can be taken as $\lambda\omega_1 \cong \frac{2}{3}(0,\tau,-\sigma)$, $\lambda r_1\omega_1 \cong \frac{2}{3}(0,-\tau,\sigma)$, $\omega_4 \cong (-2,0,0)$, $r_3 r_4 \omega_4 \cong (1,-\tau,\sigma)$, and $r_4 r_3 \omega_4 \cong (1,\tau,-\sigma)$. It is plotted in Fig. 3.

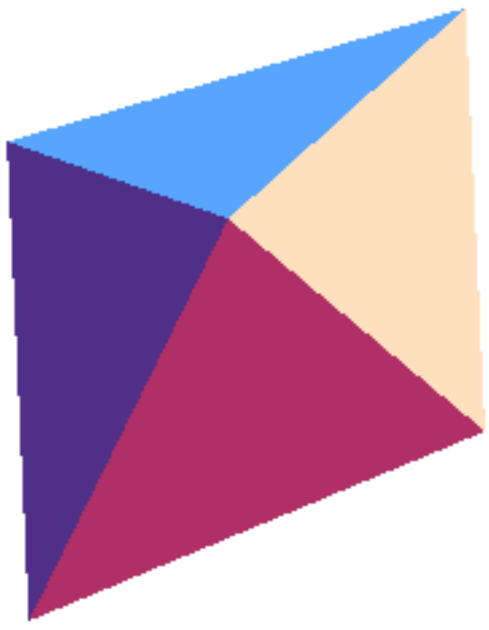

**Fig. 3.** The dipyramid, a typical cell of the dual polytope of the rectified 5-cell.

### 4.3 The dual of the polytope $O(1100) = O(\omega_1 + \omega_2)$ (truncated 5-cell)

The truncated 5-cell has 20 vertices and 10 cells consisting of 5 tetrahedra and 5 truncated tetrahedra. One tetrahedron and 3 truncated tetrahedra join to the vertex $\omega_1 + \omega_2$. The center of the tetrahedron can be represented by the vector $\lambda\omega_1$ and the centers of 3 truncated tetrahedra can be taken as $\omega_4, r_3 r_4 \omega_4, (r_3 r_4)^2 \omega_4$. Since these vectors define the hyperplane orthogonal to the vector $\omega_1 + \omega_2$ the scale factor is determined to be $\lambda = \frac{3}{7}$. Then the vertices of the dual polytope of the truncated 5-cell consist of the union of the orbits $\frac{3}{7}O(1000) \oplus (0001)$. They lie on two concentric 3-spheres with the ratio of the radii $\frac{3}{7}$. Three of the points on the outer sphere forming an equilateral triangle are connected to one point in the inner sphere forming a pyramid with edge lengths 1.41 and 0.88 as shown in Fig. 4. The base of the pyramid is an equilateral triangle with edge 1.41 and the sides are isosceles triangles with two edges equal to 0.88. The dual polytope has 10 vertices, 30 edges, 40 faces and 20 cells.

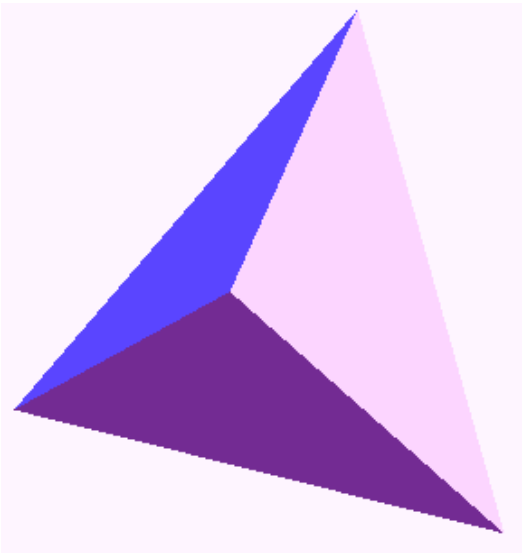

**Fig. 4.** The pyramid with equilateral triangle base

### 4.4 The dual of the polytope $O(1010) = O(\omega_1 + \omega_3)$ (cantellated 5-cell)

The symmetry leaving the vector $\omega_1 + \omega_3$ is the group generated by $r_2$ and $r_4$ of order 4 which shows that the cantellated 5-cell has 30 vertices. It has three types of cells; 5 octahedral cells, 5 cuboctahedral cells and 10 triangular prismatic cells. Of these cells 1 octahedral cell, 2 cuboctahedral cells and 2 prismatic cells meet at the vertex $\omega_1 + \omega_3$. The centers of these cells up to some scale vectors are represented respectively by the vectors $\lambda\omega_1, \rho\omega_4, \rho r_4 \omega_4$ and $\omega_2, r_2\omega_2$. The parameters are determined to be $\lambda = \frac{7}{6}$ and $\rho = \frac{7}{4}$. Vertices of the dual polytope lie on the union of the orbits $\frac{7}{6}O(1000) \oplus (0100) \oplus \frac{7}{4}O(0001)$ which determine three concentric 3-spheres with respective radii 1.044, 1.095 and 1.565. The typical cell of the dual polytope is a solid with five vertices shown in Fig. 5. The coordinates in 3D can be written as $\lambda\omega_1 \cong \frac{1}{6}(\sigma,\tau,-1)$, $\rho\omega_4 \cong \frac{1}{2}(-1,-1,0)$, $\rho r_4 \omega_4 \cong \frac{1}{4}(\tau^2,\sigma^2,1)$, $\omega_2 \cong \frac{1}{7}(1,-1,-\sqrt{5})$ and $r_2\omega_2 \cong \frac{1}{7}(-1,1,\sqrt{5})$.

Each cell of the dual polytope has the symmetry represented by the Klein 4-group $C_2 \times C_2$ generated by the $r_2$ and $r_4$.

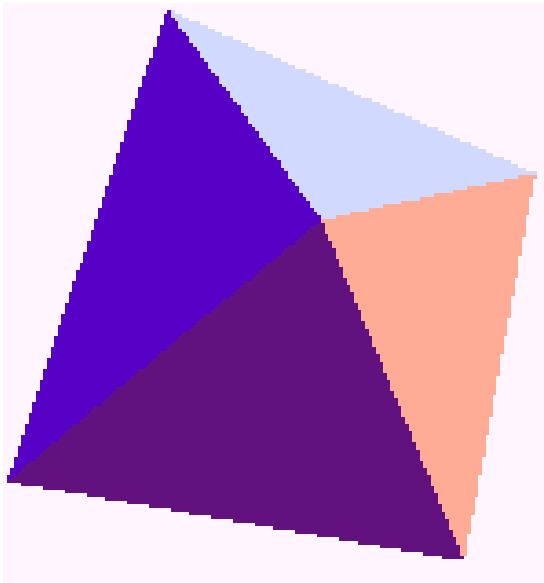

**Fig. 5.** The typical cell of the dual polytope of the cantellated 5-cell

**4.5 The dual of the polytope $O(1001) = O(\omega_1 + \omega_4)$ (runcinated 5-cell)**

The vectors in the orbit $O(1001) = O(\omega_1 + \omega_4)$ represent the 20 non-zero weights of the adjoint representation of the $SU(5)$ Lie algebra. The runcinated 5-cell has the larger symmetry $Aut(A_4)$ of order 240. The generators $r_2$ and $r_3$ fixing the vector $\omega_1 + \omega_4$ form a dihedral group of order 6 indicating that the polytope has 20 vertices. The polytope has 30 cells; 10 of which are the tetrahedral cells and the remaining 20 cells are the triangular prisms. Each type of cells is also divided into two classes. 5 tetrahedral cells are obtained by the group acting on the vector $\omega_1$ while the other 5 are obtained from the vector $\omega_4$. There is only 1 tetrahedron of each type joining to the vertex $\omega_1 + \omega_4$. Therefore the centers of two tetrahedra sharing the vector $\omega_1 + \omega_4$ can be taken as $\omega_1$ and $\omega_4$. There are 6 triangular prisms meeting at the vertex $\omega_1 + \omega_4$ the centers of the 3 prisms are represented by the vectors $\omega_3, r_2r_3\omega_3, (r_2r_3)^2\omega_3$ permuted by the dihedral symmetry $D_3 = \langle r_2, r_3 \rangle$ and the centers of the other 3 triangular prisms are represented by the vectors $\omega_2, r_2r_3\omega_2, (r_2r_3)^2\omega_2$ which are also permuted by the same dihedral symmetry $D_3$. Note that the vectors $\omega_1$ and $\omega_4$ are invariant vectors under the dihedral symmetry $D_3$. It is then clear that the union of the orbits $O(1000) \oplus (0100) \oplus O(0010) \oplus O(0001)$ constitute 30 vertices of the dual polytope of the runcinated 5-cell which has 20 cells whose centers are determined by the vectors of the orbit $O(1001) = O(\omega_1 + \omega_4)$. It follows from (15) that the norms of the vectors satisfy $|\omega_1| = |\omega_4| = \frac{2}{\sqrt{5}}$ and $|\omega_2| = |\omega_3| = \sqrt{\frac{6}{5}}$. Therefore 10 of the vertices lie on a 3-sphere with the radius $\frac{2}{\sqrt{5}}$ and the other 20 vertices lie on a 3-sphere with radius $\sqrt{\frac{6}{5}}$. Since the 8 vertices are in the intersections of the 3-spheres and the hyperplane orthogonal to the vector $\omega_1+\omega_4$ we have to express the vectors in (16) in a new basis defined by $q_0=\frac{\omega_1+\omega_4}{|\omega_1+\omega_4|}, q_1=e_1 q_0, q_2=e_2 q_0, q_3=e_3 q_0$. These 8 vectors in terms of the new basis read

$$
\begin{aligned}
&\omega_1=\frac{q_0}{\sqrt{2}}+\frac{1}{\sqrt{10}}(q_1+q_2-q_3), && \omega_4=\frac{q_0}{\sqrt{2}}-\frac{1}{\sqrt{10}}(q_1+q_2-q_3)\\
&\omega_3=\frac{q_0}{\sqrt{2}}+\frac{1}{\sqrt{10}}(\tau q_1+\sigma q_2+2q_3), && r_2r_3\omega_3=\frac{q_0}{\sqrt{2}}+\frac{1}{\sqrt{10}}(\sigma q_1-2q_2-\tau q_3),\\
&r_3r_2\omega_3=\frac{q_0}{\sqrt{2}}+\frac{1}{\sqrt{10}}(-2q_1+\tau q_2-\sigma q_3), && \omega_2=\frac{q_0}{\sqrt{2}}+\frac{1}{\sqrt{10}}(2q_1+-\tau q_2+\sigma q_3),\\
&r_2r_3\omega_2=\frac{q_0}{\sqrt{2}}+\frac{1}{\sqrt{10}}(-\sigma q_1+2q_2+\tau q_3), && r_3r_2\omega_2=\frac{q_0}{\sqrt{2}}+\frac{1}{\sqrt{10}}(-\tau q_1-\sigma q_2-2q_3).
\end{aligned}
\tag{36}
$$

All 8 vectors have the same $q_0$ component as expected. Now, one can plot the cell in terms of the components of the unit quaternions $q_1, q_2$ and $q_3$. It is a rhombohedra with $D_3$ symmetry as shown in Fig.6.

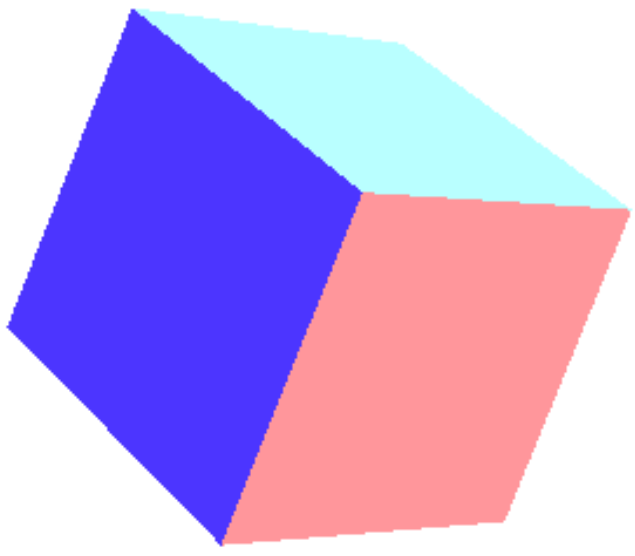

**Fig. 6.** A rhombohedron with $D_3$ symmetry.

### 4.6 The dual of the polytope $O(0110)=O(\omega_2+\omega_3)$ (bitruncated 5-cell)

The symmetry of the polytope bitruncated 5-cell is the group $Aut(A_4)$. As we noted in Section 3 it has 10 cells consisting of truncated tetrahedra, half is the dual of the other half. Four truncated cells meet at the vertex $\omega_2+\omega_3$ whose centers can be represented by the vectors $\lambda\omega_1, r_1\lambda\omega_1, \omega_4$ and $r_4\omega_4$. These vectors respectively are left invariant by the respective tetrahedral subgroups

$$\langle r_2,r_3,r_4\rangle,\ \langle r_3,r_4,r_0\rangle,\langle r_1,r_2,r_3\rangle \text{and} \langle r_0,r_1,r_2\rangle .$$

The scale factor can be determined as $\lambda = 1$. The vertices of the dual polytope lie on the union of the orbits $O(1000) \oplus O(0001)$. The vertices of the dual polytope form a 3-sphere with the radius $\frac{2}{\sqrt{5}}$. The dual polytope here is both cell transitive as well as vertex transitive. Similar to what we have explained in Section 4.5 the four vertices $\omega_1, r_1\omega_1, \omega_4, r_4\omega_4$ form the vertices of the cell of the dual polytope and they can be expressed in a basis where the unit vectors can be taken as $q_0 = \frac{\omega_2 + \omega_3}{|\omega_2 + \omega_3|}, q_1 = e_1 q_0, q_2 = e_2 q_0, q_3 = e_3 q_0$. Since all four vectors will have the same $q_0$ component we will simply express four vectors in terms of the other unit vectors. Then they are, besides an overall scale factor, given by the vectors

$\omega_1 \cong (\sigma, 2\tau, -\sigma^2), r_1\omega_1 \cong (\tau, -\tau^2, 2\sigma), \omega_4 \cong (-3, -1, -1), r_4\omega_4 \cong (2, \sigma^2, \tau^2)$.

The four vertices constitute a solid with Klein 4- group symmetry $C_2 \times C_2$ generated by the generators $r_1$ and $r_4$. The cell has four faces, all equal each other, made of isosceles triangles of sides equal to $\sqrt{\frac{6}{5}}, \sqrt{\frac{6}{5}}$ and $\sqrt{2}$ units.

The cell of the dual polytope is shown in Fig. 7.

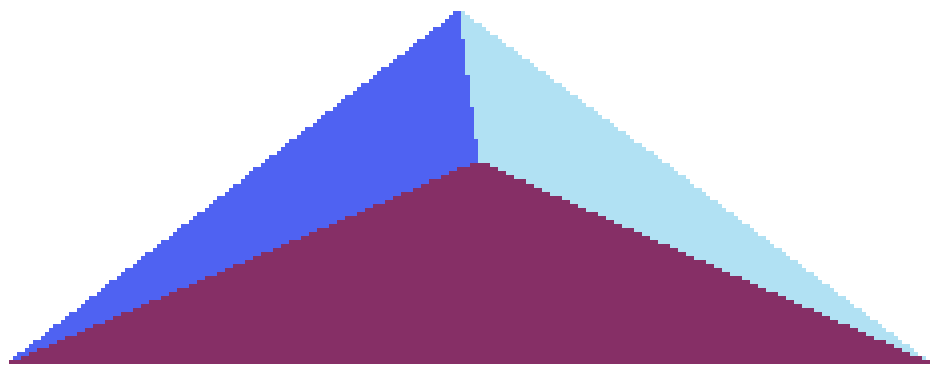

**Fig. 7.** A typical cell of the dual polytope of the bitruncated 5-cell.

### 4.7 The dual of the polytope $O(1110) = O(\omega_1 + \omega_2 + \omega_3)$ ( cantitrancated 5-cell )

Since the polytope cantitrancated 5-cell has 5 truncated octahedra, 5 truncated tetrahedra and 10 triangular prisms as cells then the dual polytope will have 20 vertices. To obtain the set of vertices we determine the centers of the cells joining to the vector $\omega_1 + \omega_2 + \omega_3$. There are two truncated octahedra with the centers represented by the vectors $\omega_4$ and $r_4\omega_4$, one truncated tetrahedron with the center represented by the vector $\lambda\omega_1$ and one triangular prism with a center represented by $\rho\omega_2$. The scale factors can be determined as $\lambda = \frac{2}{3}$ and $\rho = \frac{6}{13}$. Vertices of the dual cell lie on the union of the orbits $\frac{2}{3}O(1000) \oplus \frac{6}{13}O(0100) \oplus O(0001)$. The coordinates of a typical cell in 3D are given by the set of vectors

$$\lambda\omega_1 \cong \frac{2}{3}(\sigma,-\sigma,-\sigma^2),\ \rho\omega_2 \cong \frac{6}{13}(1,-1,-\sqrt{5}),\ \omega_4 \cong (-4,-2,0),\ r_4\omega_4 \cong (\tau+2,\sigma^3,\tau^2).$$

The cell is plotted in Fig. 8. The cell has a reflection symmetry represented by the generator $r_4$. The dual polytope consists of 60 cells of this type.

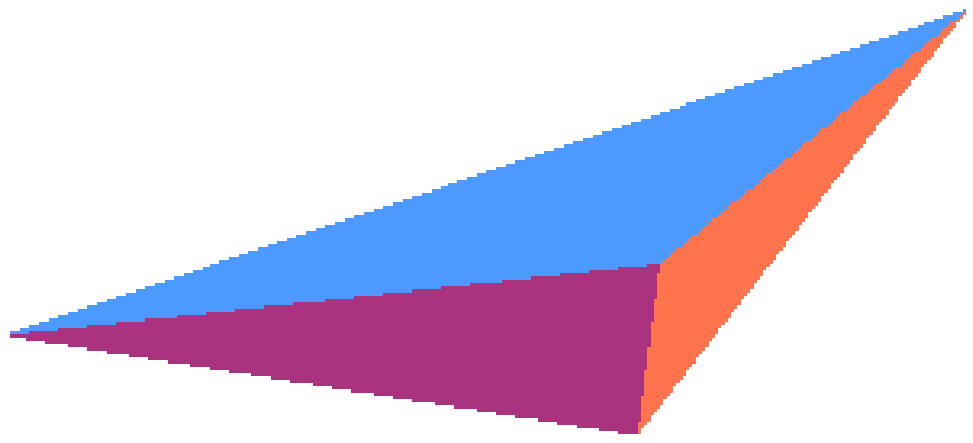

**Fig. 8.** A cell of the dual polytope of the cantitrancated 5-cell.

### 4.8 Dual of the polytope $O(1101) = O(\omega_1 + \omega_2 + \omega_4)$ (runcitruncated 5-cell)

There are five cells joined to the vector $\omega_1 + \omega_2 + \omega_4$; one is a truncated tetrahedron with a center $\lambda\omega_4$, the other one is a cuboctahedron with the center $\rho\omega_1$ and two hexagonal prisms with the centers $\omega_3$ and $r_3\omega_3$. The center of the triangular prism can be taken as $\eta\omega_2$. Then the union of the orbits $\frac{9}{8}O(1000) \oplus \frac{9}{11}O(0100) \oplus O(0010) \oplus \frac{9}{7}O(0001)$ constitutes the 30 vertices of the dual polytope. The coordinates of a typical cell in 3D are given by the set of vectors

$$\lambda\omega_4 \cong \frac{9}{7}(-3,-1,1),\ \rho\omega_1 \cong \frac{9}{8}(1,\tau^2,-\sigma^2),\ \omega_3 \cong (-\sigma,2\sigma,\tau+2),\ r_3\omega_3 \cong (-\tau,-(\sigma+2),-2\tau)$$

$$\eta\omega_1 \cong \frac{9}{11}(2,-\tau,\sigma).$$

The cell is plotted in Fig. 9. This cell has reflection symmetry under the generator $r_3$. The dual polytope consists of 60 cells of this type.

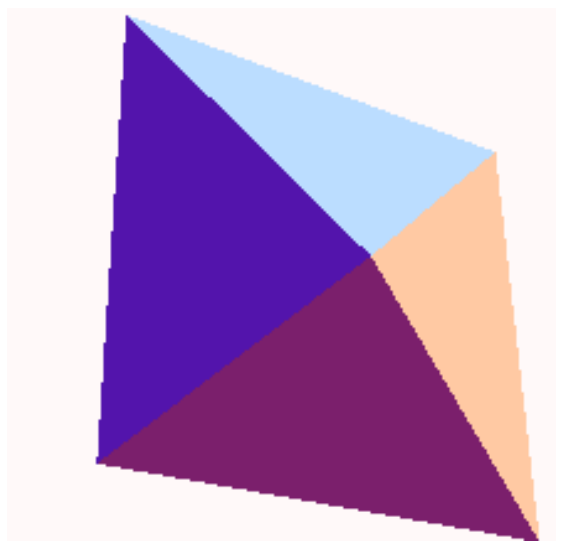

**Fig. 9.** A cell of the dual polytope of the runcitruncated 5-cell $O(1101)$.

**4.9 Dual of the polytope $O(1011)=O(\omega_1+\omega_3+\omega_4)$ (runcitruncated 5-cell)**

The dual polytope is obtained from the dual of the polytope $O(1101)=O(\omega_1+\omega_2+\omega_4)$ by the Dynkin diagram symmetry $\omega_1 \leftrightarrow \omega_4$ and $\omega_2 \leftrightarrow \omega_3$ . Then the union of the orbits $\frac{9}{8}O(0001)\oplus O(0100)\oplus\frac{9}{11}(0010)\oplus\frac{9}{7}O(1000)$ constitutes the 30 vertices of the dual polytope. The coordinates of a typical cell in 3D are given by the set of vectors

$\frac{9}{7}\omega_1 \cong \frac{9}{7}(\sigma^2,\tau^2,-2), \frac{9}{8}\omega_4 \cong \frac{9}{4}(-1,-1,0), \omega_2 \cong (1-3,-\sqrt{5}), r_2\omega_2 \cong (-1,3,\sqrt{5}), \frac{9}{11}(\tau,\sigma,2)$.

The cell is plotted in Fig. 10. This cell has reflection symmetry under the generator $r_2$. Over all the dual polytope has 60 cells.

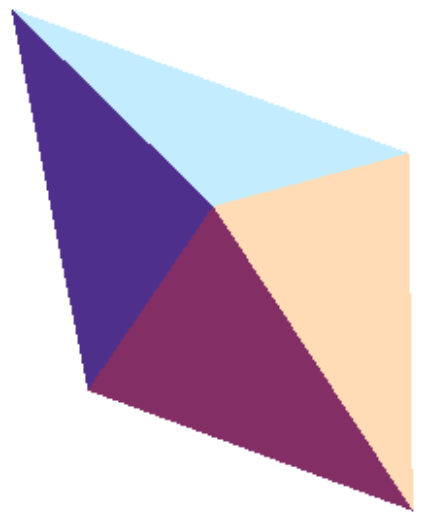

**Fig. 10.** A cell of the dual polytope of the runcitruncated 5-cell $O(1011)$.

**4.10 Dual of the polytope $O(1111)=O(\omega_1+\omega_2+\omega_3+\omega_4)$ (omnitruncated 5-cell)**

The dual polytope has 30 vertices and 120 cells. There are four cells meeting at the vertex $\omega_1+\omega_2+\omega_3+\omega_4$. Two are truncated tetrahedra with the centers represented by the vectors $\omega_1$ and $\omega_4$ up to some scale vectors. The other two cells are hexagonal prisms with the centers $\omega_2$ and $\omega_3$ up to some scale factors.

Let the centers be written as $\lambda\omega_1, \rho\omega_2, \eta\omega_3$ and $\omega_4$. Then the union of the orbits $O(1000)\oplus\frac{2}{3}O(0100)\oplus\frac{2}{3}O(0010)\oplus O(0001)$ represents the 30 vertices of the dual polytope. The projections of the four vectors $\omega_1, \frac{2}{3}\omega_2, \frac{2}{3}\omega_3$ and $\omega_4$ into three dimensions which result as the intersection of the 3-sphere of the vertices $O(1111)$ and the hyperplane orthogonal to the vector $\omega_1+\omega_2+\omega_3+\omega_4$ will represent a typical cell of the dual polytope. The coordinates of a typical cell in 3D are given by the set of vectors

$$\omega_1 \cong (\sigma^2, \tau^3, -(\sigma+2)), \frac{2}{3}\omega_2 \cong \frac{2}{3}(3, -1, -\sqrt{5}), \frac{2}{3}\omega_3 \cong \frac{2}{3}(-\sigma, 2\sigma, \tau+2), \omega_4 \cong (-4, -2, 0).$$

The cell is plotted in Fig. 11. It reflects the Dynkin diagram symmetry. The solid is closed by four planes made of two scalene triangles. The planes (123) and (234) are identical. Similarly the planes (124) and (134) are identical so the Dynkin diagram symmetry interchanges identical planes of the solid. The dual polytope has 120 cells.

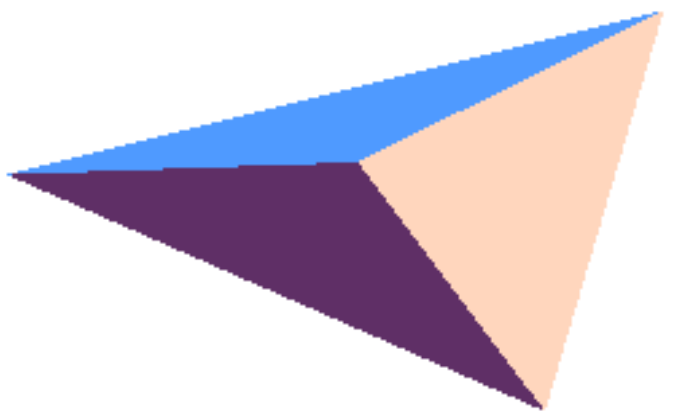

**Fig. 11.** A cell of the dual polytope of the omnitruncated 5-cell.

## 5. Conclusion

4D polytopes can be classified with respect to their symmetries represented by the Coxeter groups $W(A_4)$, $W(B_4)$, $W(H_4)$ and $W(F_4)$; the regular and semi regular polytopes are all well known in the mathematical literature. However their constructions in terms of quaternions are recently gaining some interest. Moreover their duals have not been constructed to date. In this paper we have first constructed the group $W(A_4)$ in terms of quaternions and shown its relation to the binary icosahedral group represented in two different forms by the sets of quaternions $I$ and $\tilde{I} = I(\tau \leftrightarrow \sigma)$. We have constructed semi regular polytopes in terms of quaternions and projected them to 3D by decomposing the $W(A_4)$ orbits under the Coxeter-Weyl group $W(A_3)$. The vertices of the dual polytopes of the semi regular $W(A_4)$ polytopes have been constructed and their cell structures have been identified. Since the Lie group $SU(5)$ associated with the Coxeter-Weyl group $W(A_4)$ seems to play some role in the High Energy Physics the topic that we have studied in this paper may have some impact on the problems of High Energy Physics. We shall continue to study the dual polytopes of the semi regular polytopes described by the Coxeter groups $W(B_4)$, $W(H_4)$ and $W(F_4)$ [19-20].